# Lensless polarimetric coded ptychography for high-resolution, high-throughput gigapixel birefringence imaging on a chip

Liming Yang[1,2], Ruihai Wang[2], Qianhao Zhao[2], Pengming Song[2], Shaowei Jinag[2], Tianbo Wang[2], Xiaopeng Shao[1], Chengfei Guo[2], Rishikesh Pandey[2], and Guoan Zheng[2,*]

[1]School of Optoelectronic Engineering, Xidian University, Xi'an, 710071, China
[2]Department of Biomedical Engineering, University of Connecticut, Storrs, CT 06269, USA

*Correspondence: G. Z. (guoan.zheng@uconn.edu)

**Abstract:** Polarimetric imaging provides valuable insights into the polarization state of light interacting with a sample. It can infer crucial birefringence properties of bio-specimens without using any labels, thereby facilitating the diagnosis of diseases such as cancer and osteoarthritis. In this study, we present a novel polarimetric coded ptychography (pol-CP) approach that enables high-resolution, high-throughput gigapixel birefringence imaging on a chip. Our platform deviates from traditional lens-based polarization systems by employing an integrated polarimetric coded sensor for lensless coherent diffraction imaging. Utilizing Jones calculus, we quantitatively determine the birefringence retardance and orientation information of bio-specimens from the recovered images. Our portable pol-CP prototype can resolve the 435-nm linewidth on the resolution target and the imaging field of view for a single acquisition is limited only by the detector size of 41 mm². The prototype allows for the acquisition of gigapixel birefringence images with a 180-mm² field of view in ~3.5 minutes, a performance that rivals high-end whole slide scanner but a small fraction of the cost. To demonstrate its biomedical applications, we perform high-throughput imaging of malaria-infected blood smears, locating parasites using birefringence contrast. We also generate birefringence maps of label-free thyroid smears to identify thyroid follicles. Notably, the recovered birefringence maps emphasize the same regions as autofluorescence images, underscoring the potential for rapid on-site evaluation of label-free biopsies. Our approach provides a turnkey and portable solution for lensless polarimetric analysis on a chip, with promising applications in disease diagnosis, crystal screening, and label-free chemical imaging, particularly in resource-constrained environments.

# 1. Introduction

Polarimetric imaging has attracted considerable interest in the field of biomedicine, owing to its ability to unveil structural and chemical information that often remains undetected by conventional imaging techniques [1, 2]. A notable benefit of polarimetric imaging is its application in label-free biospecimen characterization. By analyzing the polarization properties of light interacting with biospecimens, microscopic-level information about the composition and structure can be obtained without requiring labeling agents. For example, collagen fibers in connective tissues can alter the polarization state of light passing through them. Measuring this alteration provides information about collagen fiber orientation and density, which are associated with various diseases such as cancer and fibrosis [3, 4]. Another advantage of polarimetric imaging is image contrast enhancement. By selectively filtering out light waves with certain polarization states, it is possible to enhance image contrast and improve the visibility of certain structures or components based on the birefringence maps. For example, birefringence maps have been used to locate and identify malaria hemozoin crystals generated from the hemoglobin metabolism process of malaria parasites [5]. Similarly, birefringence maps have shown clinical value in screening and imaging meiotic spindles [6], monosodium urate [7], squamous cell carcinoma [8, 9], articular cartilage [10], cerebral amyloid [11], normal versus cancer cells [12], among others. These diverse applications demonstrate the potential of polarimetric imaging in enhancing diagnostic accuracy and facilitating the early detection of various medical conditions.

Conventional lens-based polarization light microscopy utilizes a polarizer-analyzer pair to measure the changes in the polarization state of light induced by the sample. Despite the wide-spread use in different application settings, there are several limitations of these lens-based systems. One significant drawback is the trade-off between the spatial resolution and the imaging field of view. One can get high resolution with a small field of view or low resolution with a large field of view, but not both [13, 14]. To obtain a high-resolution whole slide image of the sample, integration of the polarization optics with a whole slide imaging system is necessary for data acquisition. However, this integration remains elusive as whole slide scanners are often dedicated and costly tools [15, 16]. Additionally, high numerical aperture (NA) objective lenses imply a small depth of field, typically on the micron scale, which presents a challenge when tracking the axial topography variations of biospecimen. If the sample is not positioned within the objective lens's small depth of field, the image quality will degrade, necessitating rescanning and causing workflow delays [15]. Furthermore, the specialized and non-portable nature of whole slide scanner creates obstacles that prevents its widespread use in different application settings, particularly in resource-constrained environments.

To address these limitations, researchers have developed alternative methods for polarimetric imaging. One example is the lens-based Fourier ptychographic microscopy (FPM) [17] implemented with a pixelated polarization camera [18]. In this case, the achieved spatial resolution is no longer constrained by the NA of the objective lens but determined by the largest incident angle. Consequently, a half-pitch resolution of 0.55 µm has been demonstrated over a field of view of 3.78 mm$^2$ in a polarization-sensitive FPM system, corresponding to a moderate space bandwidth product of ~12.5 megapixels [18]. Similarly, the vectorial ptychographic imaging framework [19] can also be implemented using FPM [20], where a generator polarizer is placed at the illumination path and an analyzer is placed at the detection path. With four generator-analyzer configurations, the object's Jones matrix can be recovered from the low-resolution raw measurements [20]. While these demonstrations can improve the resolution in polarimetric imaging, the achieved imaging throughput is still orders of magnitude lower than that of a whole slide scanner [15, 21]. Moreover, FPM relies on coded illumination, and its imaging model is contingent upon how the incident beam enters the sample [14, 22]. As a result, thick specimens frequently pose a challenge to this modality, necessitating an accurate model of the light-sample interaction process [23, 24]. This same challenge also applies to lensless ptychographic implementations that utilize multi-angle illumination [25-27]. The quality of the recovered images often falls short when compared to that of conventional bright-field polarization microscopy.

Another alternative to conventional lens-based polarization light microscopy is to perform polarimetric imaging without using any lenses [27-32]. For example, multi-height phase retrieval has been implemented with a polarizer-analyzer pair for lensless polarimetric imaging, demonstrating potentials for biomedical applications [28]. However, the traditional multi-height phase retrieval technique may not provide sufficient diversity measurements for recovering complex objects with both intensity and phase properties [33]. Specifically, recovering bio-specimens containing $2\pi$ phase wraps, such as bacterial colonies or cytology smears, is a challenging task for this approach [33, 34]. This difficulty arises because the object's phase information cannot be effectively converted into intensity variations for detection—the phase transfer function approaches zero [22, 35, 36]. Moreover, existing lensless polarimetric imaging implementations still face constraints in spatial resolution, field of view, and imaging throughput when compared to robotic microscope platforms.

In this work, we present a novel polarimetric coded ptychography (pol-CP) technique that enables high-resolution, high-throughput gigapixel birefringence imaging on a chip. Instead of using conventional lens-based polarization optics, our platform adopts the concept of coded ptychography for high-resolution, high-throughput coherent diffraction imaging with a disorder-engineered surface [34, 36-40]. In our implementation, we illuminate the specimen with circularly polarized laser light and attach an ultra-thin polarizer film to a blood-coated sensor for image acquisition. The integrated coded sensor is then mounted on a compact, customized stage, enabling control over both rotational and translational motion. By analyzing four different polarization states with Jones calculus, we retrieve the quantitative birefringence retardance and orientation information of the specimen. Our cost-effective and portable pol-CP prototype can resolve the 435-nm linewidth on the resolution target, with the field of view of a single acquisition limited only by the size of the sensor -- approximately 41 mm$^2$. Utilizing this prototype platform, we demonstrate that gigapixel birefringence images with a 180-mm$^2$ field of view can be acquired in approximately 3.5 minutes, a performance that rivals high-end whole slide scanner but at a small fraction of the cost. In contrast to conventional lens-based polarization microscopy, our platform enables image refocusing after data acquisition, effectively addressing the autofocusing issue present in traditional whole slide scanners. Furthermore, the reported technique models only the complex wavefront exiting the sample, not its entry, rendering sample thickness inconsequential in the image formation process. Unlike FPM [17, 18, 20] and other ptychographic implementations [25-27], which necessitate thin sample assumptions, our platform eliminates this constraint. Given these advantages, we anticipate that this approach will find widespread applications in disease diagnosis, crystal screening, and label-free chemical imaging, particularly in resource-constrained environments.

## 2. Materials and methods

### 2.1 Polarimetric coded ptychography (pol-CP) on a chip

Figures 1a and 1b illustrate the schematic of our pol-CP prototype setup. This prototype consists of three main components (from top to bottom): the illumination light source, the integrated polarimetric coded sensor, and the motion system for programmable translation and rotation control. As shown in Fig. 1a, a fiber-coupled, 5-mW 405-nm laser diode serves as our light source. Positioned upstream of the sample, a left-hand circular polarizer converts the linearly polarized light from the laser into circularly polarized light for sample illumination. The use of laser diode in pol-CP offers superior spatial and temporal coherence compared to conventional LED-based lensless setups. It also ensures adequate illumination intensity for a short exposure time of ~1 ms for image acquisition. This allows the coded sensor operated in continuous motion without occurrence of motion blur issues.

The imaging system of pol-CP features an integrated polarimetric coded image sensor. In our implementation, we smeared a drop of goat blood on the sensor's coverglass followed by cell fixation with ethyl alcohol (Fig. 1c). We select goat blood because of its 2-3 μm red blood cell size, which is the smallest among all animals. The smearing process naturally creates a thin, uniform, and dense monolayer of blood cells on the sensor's coverglass for light wave modulation. This blood-cell layer serves as the effective ptychographic probe beam within the pol-CP setup [22]. The fabrication process of this high-performance coded surface can be completed in ~5 minutes with minimal cost, and notably it does not require the use of complex lithographic tools typically required for the construction of metasurfaces or disorder-engineered surfaces [34]. In our pol-CP prototype, we used a drop of polydimethylsiloxane (PDMS) liquid to attach a ~80-µm polarizing film to the coded sensor, as shown in Fig. 1d. The use of PDMS ensures a secure attachment between the polarizing film and the blood-cell layer, thus forming an integrated polarimetric coded sensor capable of high-resolution ptychographic imaging. As shown in Fig. 1b, when the specimen exhibits optical anisotropy, the transmitted light becomes elliptically polarized and is subsequently collected by the integrated coded sensor.

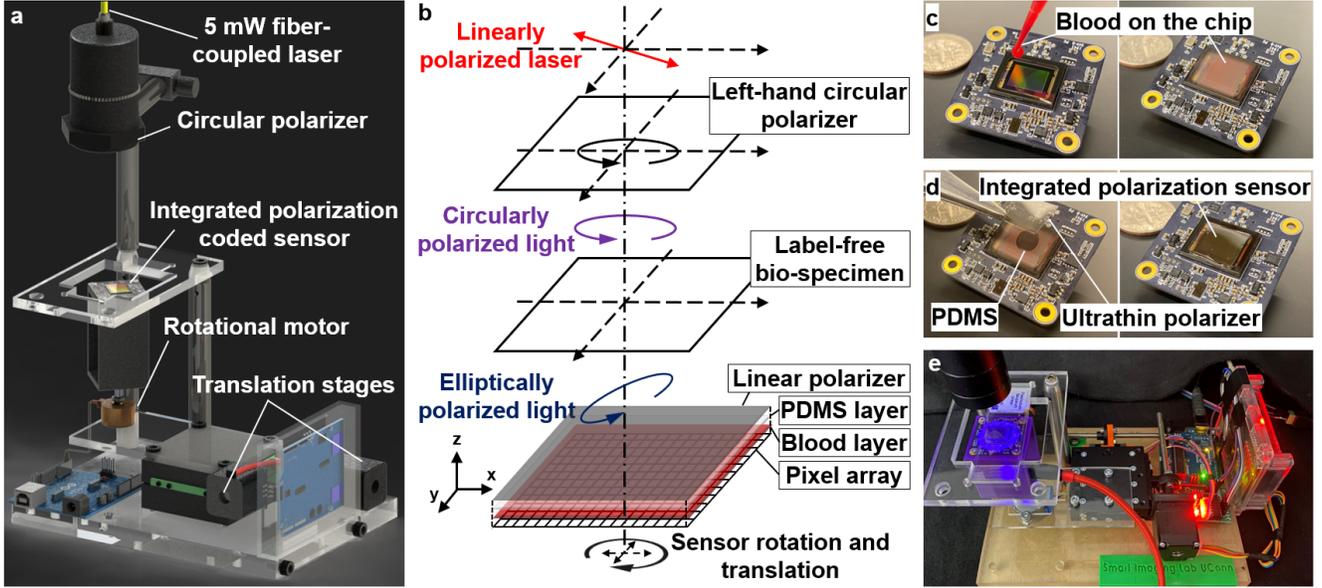

**Fig. 1.** Schematic diagram and operation of the lensless polarimetric coded ptychography (pol-CP) platform. (a) A circular polarizer transforms linearly polarized light from the laser diode into circularly polarized light, which then interacts with the specimen. The exit wave is recorded by the integrated polarimetric coded image sensor. (b) Illustration of changes in the polarization state within the pol-CP system. (c) A 2 µl sample of goat blood is smeared onto the sensor's coverglass and fixed with ethyl alcohol, forming a thin and dense blood-cell layer that acts as a high-performance scattering lens with a theoretically unlimited field of view. (d) An ultrathin polarizing film is adhered to the blood-cell layer, forming an integrated polarimetric coded image sensor. (e) The pol-CP prototype features a polarimetric coded sensor mounted on programmable stages, allowing control over both rotational and translational motions.

The motion system of our prototype device is shown in the bottom panel of Fig. 1a. We used stepper motors to build a compact motorized stage capable of managing both translational and rotational motion controls. Figure 1f shows the entire pol-CP prototype platform. In Fig. S1, we show different parts of the pol-CP setup in detail and provide the wiring diagram of the motors. The operation of the entire platform can be viewed in Visualization 1.

## 2.2 Imaging model and reconstruction

Figure 2 shows the data acquisition and reconstruction process of pol-CP. In the data acquisition process, the integrated polarimetric coded sensor is rotated to four specific angles ($\alpha = 0, 45, 90, 135°$). At each angle, a set of lensless diffraction data is obtained by translating the coded sensor to different lateral positions ($x_i, y_i$). The four sets of measurements, labelled as $I_1^\alpha \ldots I_N^\alpha$ in Fig. 2a, are subsequently utilized to recover four high-resolution images corresponding to different polarization states, as displayed in Fig. 2b. From these recovered intensity images, the optic axis orientation and retardance maps can be derived using Jones calculus, as demonstrated in Figs. 2c-2e.

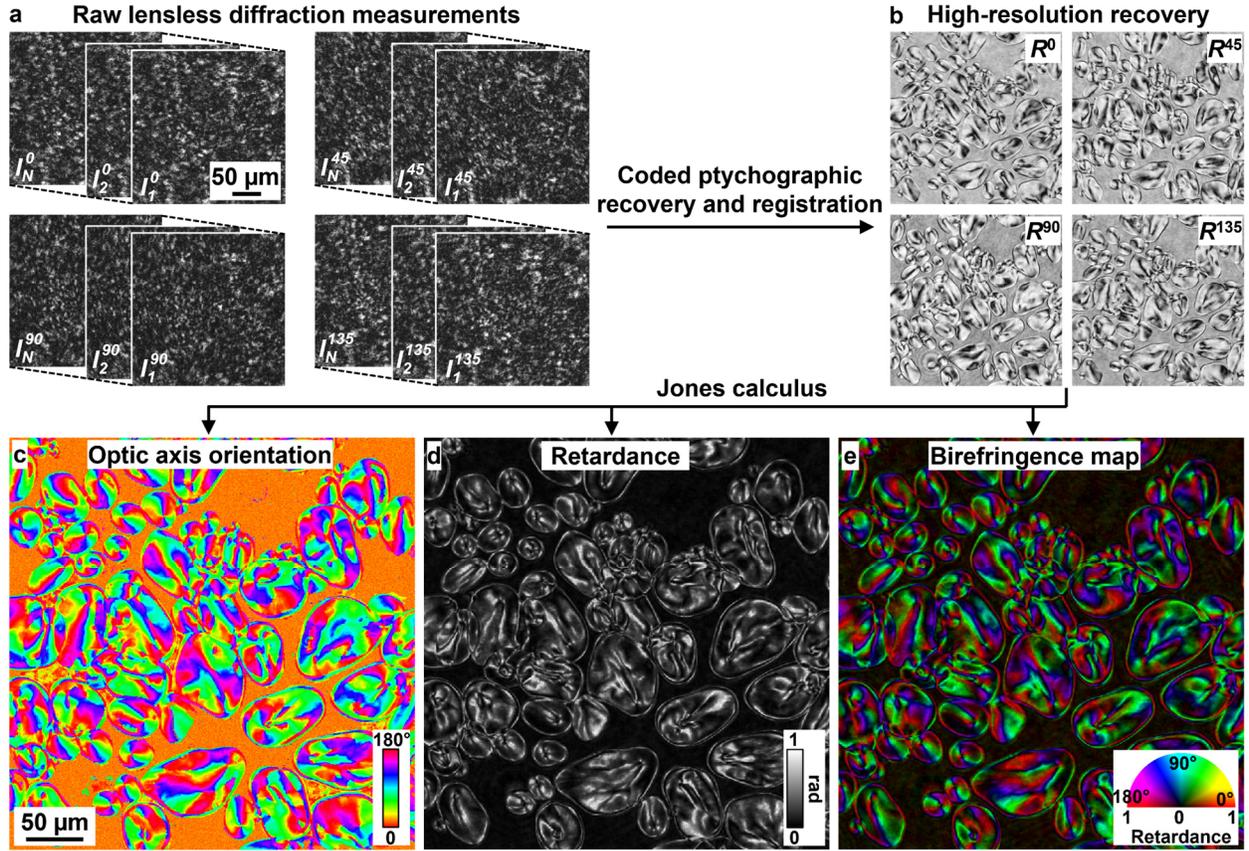

**Fig. 2.** Image acquisition and reconstruction process of pol-CP. (a) The captured four sets of raw images with the integrated coded sensor at four different orientations. (b) The recovered four images using with the 0-degree orientation as the reference. (c, d) The optic axis orientation and retardance maps are derived using Jones calculus. (e) A pseudo-coloured birefringence map by encoding the optic axis orientation as colour hue and the retardance information as intensity.

The forward imaging model of pol-CP can be discussed as follows. Initially, the input left-hand circularly polarized light interacts with the sample, producing a wave field that travels a distance $d_1$ between the sample and the integrated coded sensor. This wave field is then modulated by the linear polarizing film affixed to the integrated coded sensor. The imaging process above can be written as

$$\overrightarrow{E_{out}^{\alpha}}(x,y) = J_{pol}^{\alpha} \cdot PSF_{free}(d_1) * (J_{sample}(x,y) \cdot \overrightarrow{E_{in}}), \qquad (1)$$

where '·' represents point-wise multiplication, and '∗' represents convolution operation. In Eq. (1), $\overrightarrow{E_{in}} = \sqrt{t/2}\,[1, i]^T$, representing the Jones vector of the input circularly polarized light with intensity $t$. The term $PSF_{free}(d_1)$ represents the convolution kernel for free space propagation over a distance $d_1$. $J_{sample}(x,y)$ and $J_{pol}^{\alpha}$ in Eq. (1) correspond to the Jones matrices of the birefringent sample and the linear polarizer orientated at angle $\alpha$, respectively. In particular, the Jones matrix of the sample $J_{sample}(x,y)$ can be expressed as

$$J_{sample}(x,y) = \begin{bmatrix} \cos\left(\frac{\delta}{2}\right) + i \cdot \sin\left(\frac{\delta}{2}\right)\cos(2\theta) & i \cdot \sin\left(\frac{\delta}{2}\right)\sin(2\theta) \\ i \cdot \sin\left(\frac{\delta}{2}\right)\sin(2\theta) & \cos\left(\frac{\delta}{2}\right) - i \cdot \sin\left(\frac{\delta}{2}\right)\cos(2\theta) \end{bmatrix}, \qquad (2)$$

where $\delta(x,y)$ and $\theta(x,y)$ represent the retardance magnitude and the slow axis orientation of the birefringent sample (we ignore the $(x,y)$ coordinate for $\delta$ and $\theta$ in Eq. (2) for simplicity). $\delta(x,y)$ and $\theta(x,y)$ are the two maps that we aim to recover using the pol-CP approach. Meanwhile, the Jones matrix of the linear polarizer can be expressed as

$$J_{pol}^{\alpha} = \begin{bmatrix} \cos^2(\alpha) & \cos(\alpha)\sin(\alpha) \\ \cos(\alpha)\sin(\alpha) & \sin^2(\alpha) \end{bmatrix}, \tag{3}$$

where $\alpha$ represents the orientation angle of the linear polarizing film on the coded sensor. By rotating the entire coded sensor, we can change $\alpha$ to different angles in the image acquisition process. We also note that $\alpha$ is a constant in our implementation, and thus, $J_{pol}^{\alpha}$ does not vary across different $(x, y)$ positions. This is different from the use of commercially available polarized sensors (such as Sony IMX 250MZR), which incorporates polarizers with four different angles $\alpha$ above different pixels of the sensor.

In the second step of the imaging process, the light field emerging from Eq. (1) is further modulated by the blood-cell layer at the coded surface plane. The resultant light field then travels a distance $d_2$ between the coded surface and the pixel array. The captured intensity image can be expressed as

$$I_i^{\alpha}(x - x_i, y - y_i) = \left| \overrightarrow{E_{out}^{\alpha}}(x, y) \cdot CS(x - x_i, y - y_i) * PSF_{free}(d_2) \right|^2_{\downarrow M}, \tag{4}$$

where $I_i^{\alpha}$ represents the $i^{th}$ captured image corresponding to the positional shift of $(x_i, y_i)$, the superscript $\alpha$ of $I_i^{\alpha}$ represents the orientation angle of the integrated coded sensor, $CS(x, y)$ represents the transmission profile of the coded surface, and '$\downarrow M$' represents M-fold down-sampling by the pixel array.

The reconstruction process of pol-CP aims to recover the retardance magnitude $\delta(x, y)$ and the slow axis orientation $\theta(x, y)$ in Eq. (2), leveraging the four sets of intensity measurements $I_i^{\alpha}$ at orientation angles $\alpha = 0$, 45, 90, 135°. The detailed recovery process is provided in Supplement 1 Note 1. In this process, we first recover the lateral positional shift $(x_i, y_i)$ directly from the captured raw images [41]. We then recover the scalar wavefront $W^{\alpha}(x, y)$ at the coded surface plane following a coded ptychographic retrieval process [42]:

$$W^{\alpha}(x, y) = \left| \overrightarrow{E_{out}^{\alpha}}(x, y) \right| \cdot e^{i\rho^{\alpha}(x, y)}, \tag{5}$$

where $e^{i\rho^{\alpha}(x, y)}$ represents the phase term of the scalar exit wavefront of the birefringent sample. With $W^{\alpha}(x, y)$, we can then digitally propagate it back to the sample plane and obtain the scalar sample wavefront $S^{\alpha}(x, y)$ as follows:

$$S^{\alpha}(x, y) = W^{\alpha}(x, y) * PSF_{free}(-d_1) = \left| J_{pol}^{\alpha} \cdot J_{sample}(x, y) \cdot \overrightarrow{E_{in}} \right| \cdot e^{i\gamma^{\alpha}(x, y)}, \tag{6}$$

where $e^{i\gamma^{\alpha}(x, y)}$ represents the phase term of the birefringent sample. Subsequently, we rotate and align $S^{\alpha}(x, y)$ with respect to $S^0(x, y)$, thus obtaining $R^{\alpha}(x, y)$ in Supplementary 1 Note 1. For image rotation, we implemented a three-step shearing process in the Fourier domain [43]. For image registration, we estimate the residual transverse shift $(\widehat{\Delta x'}, \widehat{\Delta y'})$ and rotation angle $(\widehat{\Delta \sigma'})$ via a closed-form equation detailed in Supplement 1 Note 1.

After the image registration process, we have the following expressions for $R^{\alpha}(x, y)$:

$$|R^0(x, y)|^2 = \frac{t}{2}[1 - \sin(\delta(x, y))\sin(2\theta(x, y))] \tag{7}$$

$$\left|R^{45}(x, y)\right|^2 = \frac{t}{2}[1 + \sin(\delta(x, y))\cos(2\theta(x, y))] \tag{8}$$

$$|R^{90}(x, y)|^2 = \frac{t}{2}[1 + \sin(\delta(x, y))\sin(2\theta(x, y))] \tag{9}$$

$$\left|R^{135}(x, y)\right|^2 = \frac{t}{2}[1 - \sin(\delta(x, y))\cos(2\theta(x, y))] \tag{10}$$

With Eqs. (7)-(10), we can then recover the spatial distribution of retardance magnitude and the optic axis of the birefringent sample:

$$\delta(x, y) = \sin^{-1} \sqrt{p(x,y)^2 + q(x,y)^2}, \tag{11}$$

$$\theta(x, y) = \tfrac{1}{2}\tan^{-1}\left(p(x,y)/q(x,y)\right), \tag{12}$$

where $p = \frac{|R^{90}(x,y)|^2 - |R^{0}(x,y)|^2}{|R^{90}(x,y)|^2 + |R^{0}(x,y)|^2}$, $q = \frac{|R^{45}(x,y)|^2 - |R^{135}(x,y)|^2}{|R^{45}(x,y)|^2 + |R^{135}(x,y)|^2}$.

To improve visualization of the reconstructed image, a pseudo-coloured birefringence map can be obtained by encoding the optic axis orientation as colour hue and the retardance information as intensity, as shown in Fig. 2e.

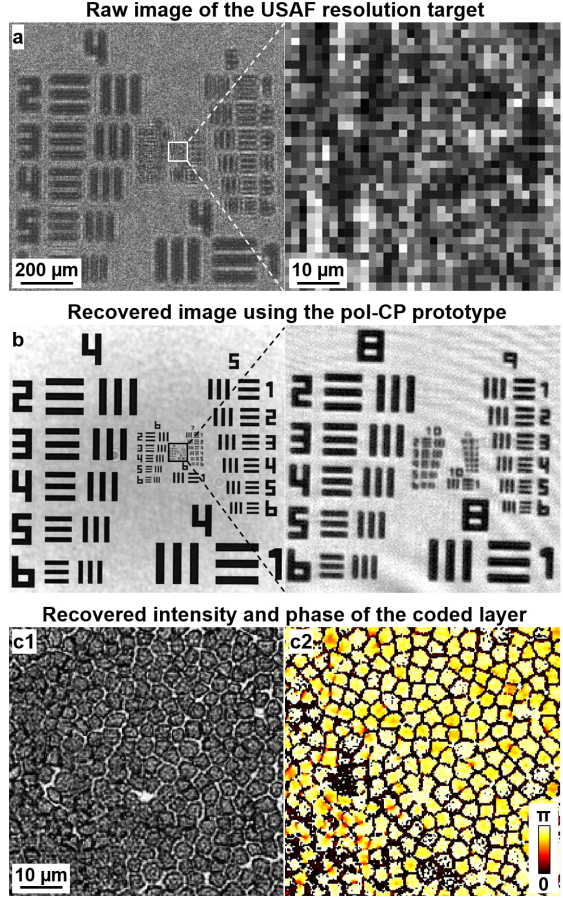

**Fig. 3.** Image quality quantification using a USAF resolution target. (a) The captured raw image of the resolution target. (b) The recovered image using the pol-CP platform. The recovered intensity (c1) and phase (c2) images of the blood-cell layer on the integrated coded sensor. The goat blood cells are smallest among all animals.

### 3. Results

#### 3.1 Imaging performance characteristics

We assessed and quantified the imaging performance of the pol-CP prototype using three types of objects: a resolution target (Fig. 3), a quantitative phase target (Fig. 4), and a potato starch granules sample (Fig. 5). A calibration experiment was designed to determine the transmission profile of the blood-coded layer on the integrated sensor. For this experiment, we selected an object devoid of slow-varying phase, typically a blood smear slide, and acquired ~1500 images by translating the coded sensor across various lateral positions. These images assisted in the recovery of both the object and the blood-coded layer on the sensor. Once the blood-cell layer is obtained, it becomes feasible to use it for subsequent experiments, reducing the number of required images to ~250 to cover one field of view (41 mm²) of the detector.

Figure 3a shows the captured raw image of the resolution target and Fig. 3b shows the recovered image using the pol-CP prototype. The 435 nm linewidth of group 10, element 2 on the target is clearly visible in the recovered image. This level of detail is particularly remarkable considering the pixel size of the detector in the captured raw image is 1.85 µm. It is worth mentioning that the achieved resolution in our demonstration of Fig. 3b ranks among the highest in lensless polarimetric imaging studies. The achieved resolution is also higher than the lens-based FPM demonstrations [18, 20] while enabling orders of magnitude larger field of view. In Fig. 3c, we present the recovered intensity and phase profiles of the blood-cell layer on the sensor, demonstrating the presence of a densely packed monolayer of goat blood cells. This blood-cell layer plays a crucial role in both intensity and phase modulation, contributing to the successful high-resolution recovery of birefringent objects [36].

To validate the quantitative nature of the pol-CP platform, we extended our evaluation of the pol-CP prototype by imaging a quantitative phase target. Figure 4a-b show the recovered amplitude and phase of the target. Figure 4c presents the recovered birefringence map of the phase target, confirming the absence of retardation within the sample.

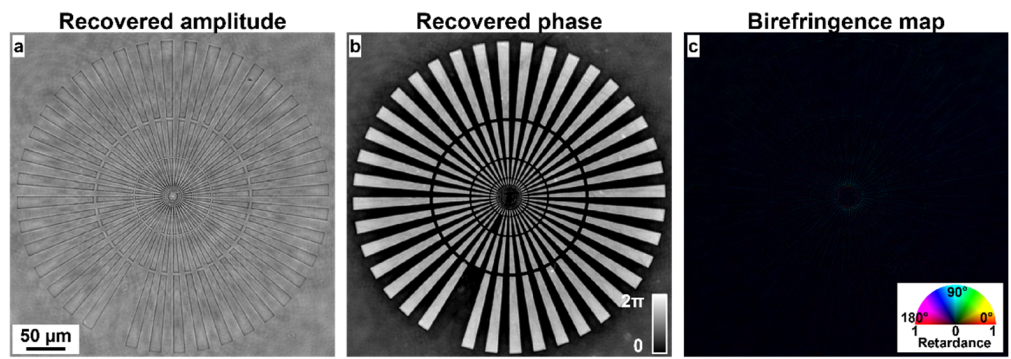

**Fig. 4.** Validating the pol-CP prototype using a quantitative phase target. The recovered amplitude (a), phase (b), and colour-coded birefringence map (c).

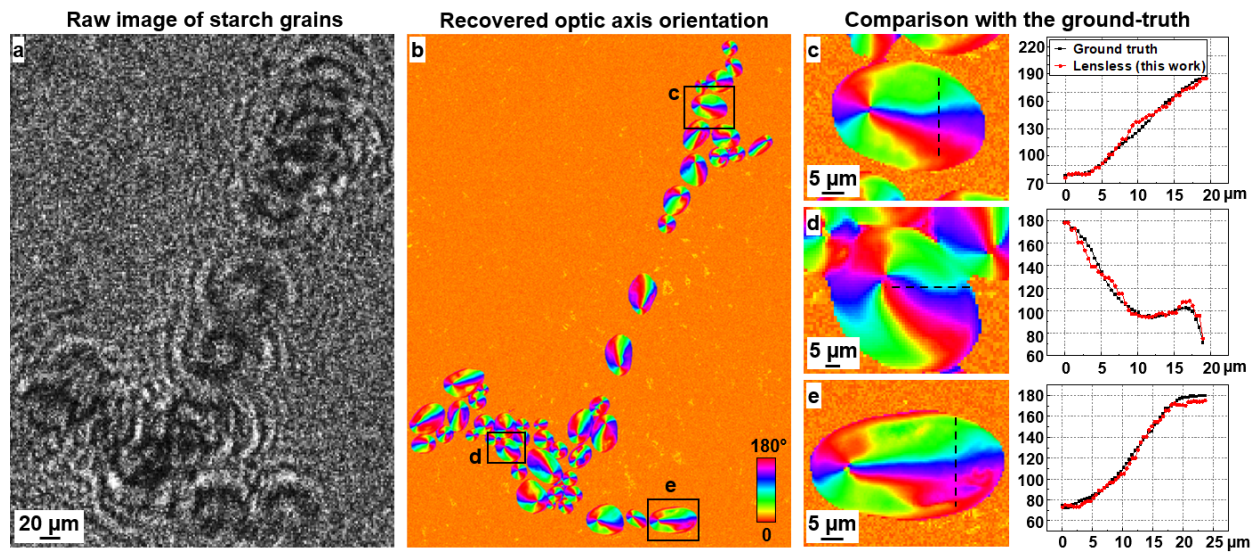

**Fig. 5.** Validation of the pol-CP prototype by imaging potato starch granules. (a) The captured raw diffraction measurement of the potato starch granules. (b) The reconstructed optic axis orientation using the pol-CP prototype. (c-e) Zoomed-in view of the highlighted regions in (b). The line traces show the measurements using the pol-CP prototype (black dotted line) and the conventional lens-based polarized light microscopy with a 10×, 0.45 NA objective (red dotted line).

In Fig. 5, we validated the pol-CP prototype with potato starch granules, which exhibit strong birefringence due to the high radial orientation of amylopectin crystallites. Figure 5a and 5b show the raw image and the reconstructed optic axis orientation of the granules using the pol-CP prototype. Figure 5c-5e provide magnified views of the three highlighted regions in Fig. 5b. To validate the accuracy of our pol-CP prototype, we also reconstructed the optic axis orientation using the gold standard lens-based polarized light microscopy (ground truth). In this experiment, a 10×, 0.45 NA was used to acquire images. The agreement between our lensless prototype (red curves) and the ground truth (black curves) in the right panels of Fig. 5c validates the effectiveness of the pol-CP approach.

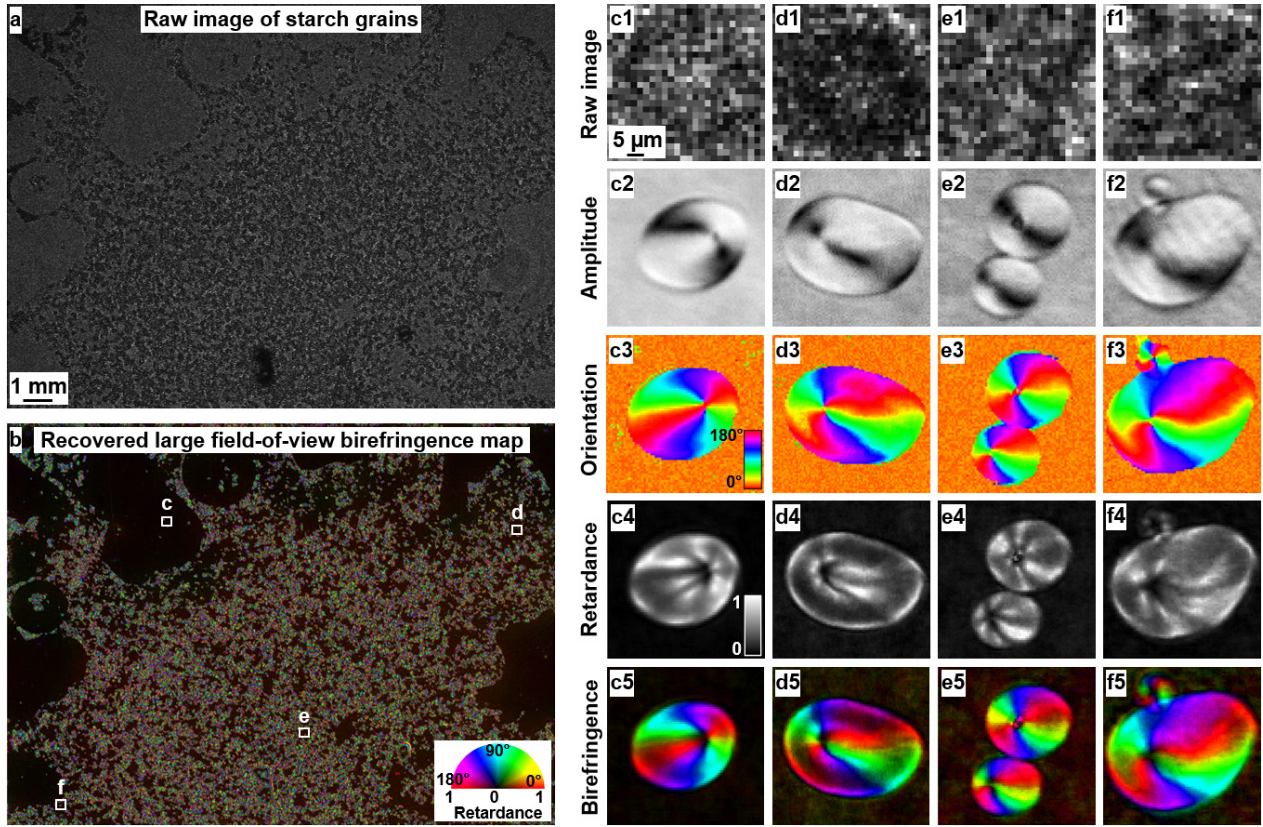

**Fig. 6.** High-resolution, large field-of-view gigapixel birefringence imaging via pol-CP. (a) The captured raw image of the whole-slide starch grain sample. (b) The recovered birefringence map. (c)-(f) The magnified views of different regions in (b). (c1)-(f1) The captured raw images. (c2-f2) The recovered amplitude. (c3)-(f3) The recovered optic axis orientation. (c4)-(f4) The recovered retardance. (c5)-(f5) The colour-coded birefringence map.

## 3.2 High-resolution gigapixel birefringence imaging over a large field-of-view

Conventional lens-based polarized light microscopy [44] has a limited space-bandwidth product that restricts its ability to deliver both high-resolution and a large field of view [13]. Our lensless pol-CP prototype offers a solution to this limitation, getting the best of both worlds. The imaging field of view of a single acquisition is limited only by the size of the image sensor. The lateral translation of the coded sensor naturally expands the field of view, enabling polarimetric whole slide imaging in high throughput. In our implementation, we typically acquire 250 images in ~9 seconds for one field of view of the detector (Fig. S2 shows the reconstructions by using different numbers of raw measurements). By translating and rotating the coded sensor to different lateral positions and orientations, we can obtain gigapixel birefringence images with a 180-mm$^2$ field of view in approximately 3.5 minutes, a performance that rivals high-end whole slide scanner but at a small fraction of the cost [15]. As an example, Figs. 6a and 6b show the raw image and recovered gigapixel birefringence map of a whole slide potato starch granule sample. Figures 6c-6f show the magnified views of four different regions, where we compare the raw measurements with the recovered amplitude, optic axis orientation, retardance, and colour-coded birefringence map.

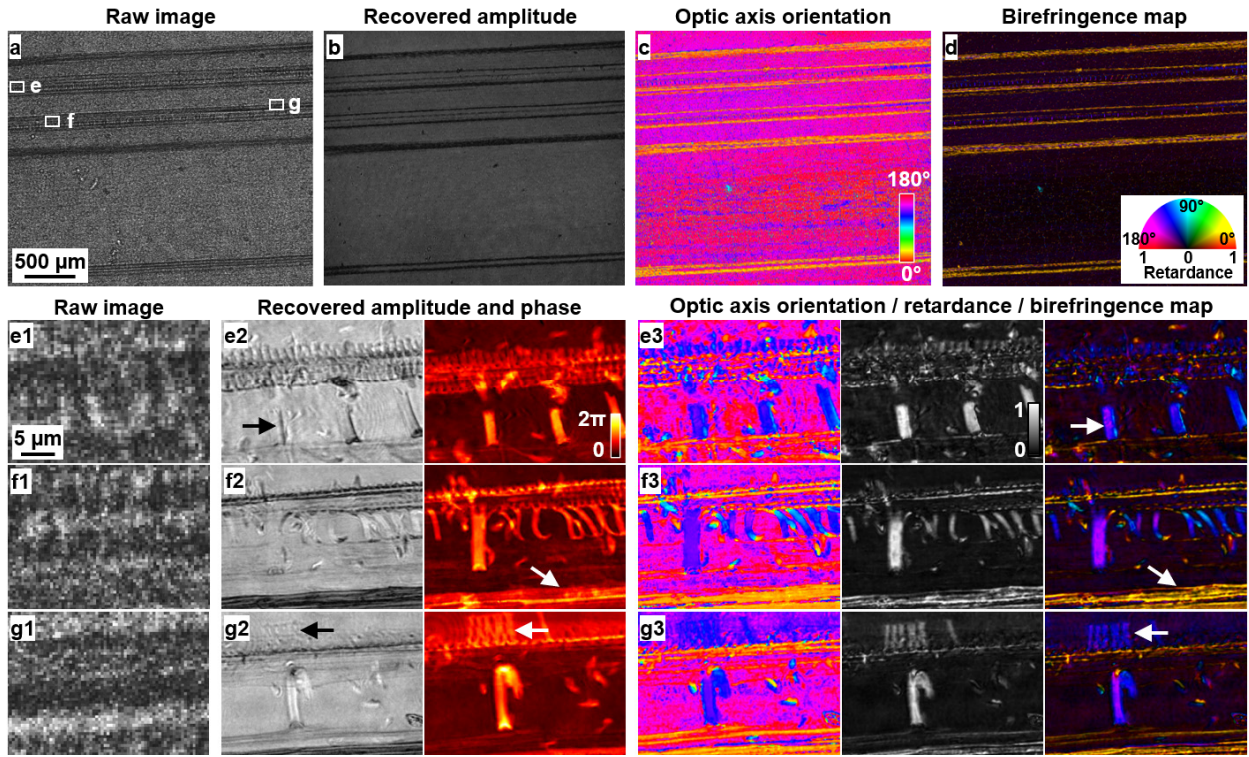

**Fig. 7.** High contrast background-free imaging of a corn stem sample. (a) Captured raw image of corn stem. (b-d) The recovered amplitude, optic axis orientation and birefringence map of (a). (e1-g3) Enlarged raw images of the highlighted regions in (a). (e1-g1) Raw images. (e2-g2) The recovered amplitude and phase. (e3-g3) The reconstructed optic axis orientation, retardance and birefringence map.

### 3.3 Imaging with quantitative polarimetric and phase contrast

Achieving good image contrast in microscopy is crucial for the study of specimen structure and enhancing diagnostic capabilities. While various methods exist to enhance image contrast in microscopy, such as fluorescence microscopy, darkfield microscopy, and differential interference contrast (DIC) microscopy, these techniques often lack the ability to provide quantitative information about the specimens without the need for external staining. For instance, fluorescence microscopy requires the application of external dyes or tags, which can potentially produce toxic by-products or lead to photobleaching. On the other hand, darkfield and DIC microscopy can enhance image contrast, but their outputs are predominantly qualitative in nature. Unlike other common microscopy modalities, the lensless pol-CP platform offers a unique solution by generating both the quantitative phase contrast and polarimetric contrast of the specimens at a high imaging throughput. The quantitative phase information is directly related to the morphology of the sample while the quantitative polarimetric contrast provides insights into the intrinsic chemical composition of the specimen.

This capability was demonstrated through the analysis of a corn stem sample in Fig. 7. In Figs 7a-7d, we show the raw image, the recovered amplitude, the recovered optic axis orientation, and the quantitative birefringence map of the corn stem. The plant cells in this specimen are composed of parallelly arranged cellulose structures, thereby exhibiting the expected birefringence property. In Figs. 7e-7g, we further show the zoomed-in views of three highlighted regions of Fig. 7a. Figures 7e1-7g1 show the captured raw diffraction patterns, and Figs. 7e2-7g2 show the recovered amplitude and phase images. These recovered images demonstrate the utility of the quantitative phase map in visualizing the otherwise transparent structure of the specimens. Figures 7e3-7g3 show the zoomed-in views of the recovered optic axis orientation, the retardance, and the colour-coded birefringence map. We can see that the orientations of the cell match well with the optic axis displayed in colour. The retardance of the cell is much higher than the background due to the different optical and structural properties. The birefringence map, which encodes both optic axis orientation and retardance information, significantly improves the contrast and makes it possible to visualize fine details of the cell that would otherwise be obscured by the background in amplitude maps. The ability to generate both quantitative phase contrast and polarimetric contrast in a single imaging technique holds great value for a wide range of applications, providing researchers with new insights into the structural and chemical properties of specimens without the need for external staining. This capability opens up new possibilities for studying a wide range of samples, from biological specimens to materials science applications, with enhanced accuracy and precision.

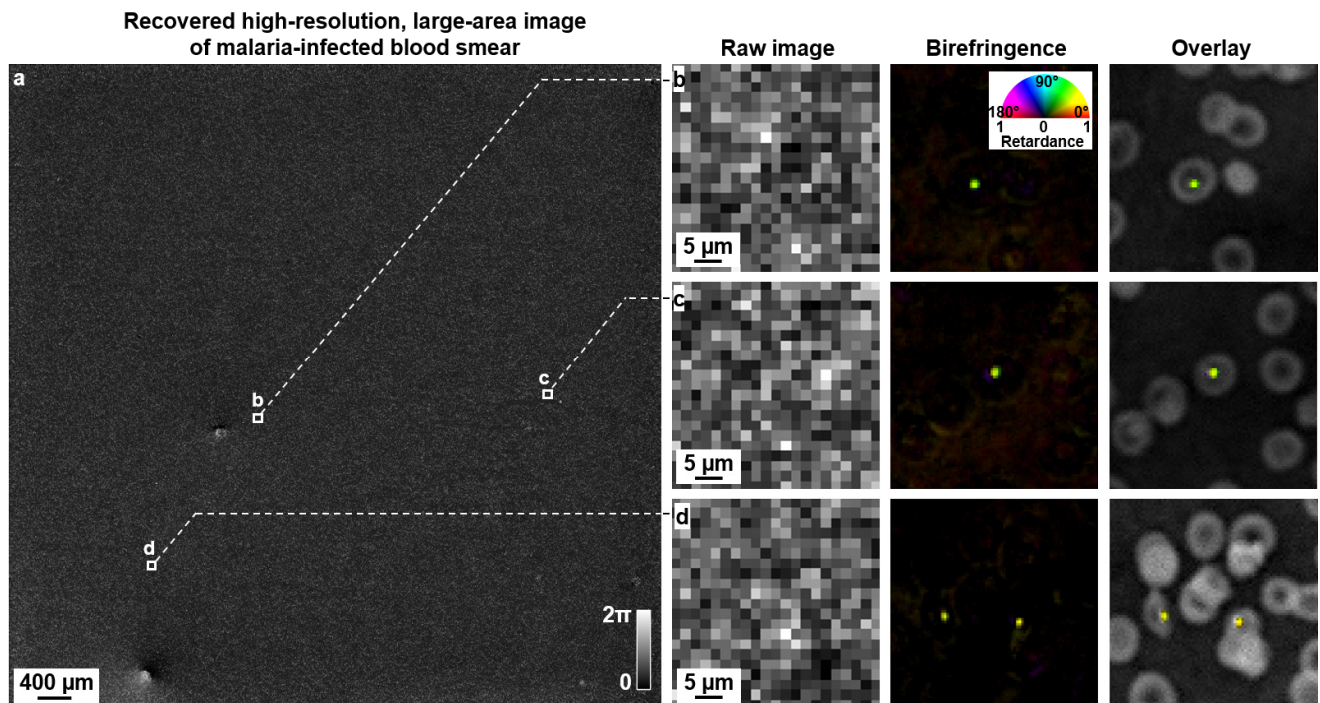

**Fig. 8.** Detecting malaria-infected blood cells using the lensless pol-CP prototype. (a) The recovered phase image of the sample. (b)-(d) The zoomed-in views of the raw diffraction measurements (left column), quantitative birefringence maps (middle column) and overlaid images of phase and birefringence map (right column).

**3.4 Tracking of malaria-infected blood cells**

The lensless pol-CP platform can be a valuable tool for malaria diagnosis in resource-limited settings. When a malarial parasite infects a red blood cell (RBC), changes occur in its hemoglobin content, affecting the birefringence property of the cell. As the parasite grows, it crystallizes hemoglobin into hemozoin, which exhibits high birefringence. Leveraging polarimetric imaging, the pol-CP platform can effectively track and detect malaria-infected RBCs by detecting the hemozoin crystals.

In Fig. 8, we analyzed the capability of the pol-CP platform for detecting hemozoin crystals on a blood smear sample infected with Plasmodium falciparum, a particularly harmful type of malaria parasite. Figure 8a shows the recovered high-resolution, large-field-of-view phase map of the blood smear slide. Three zoomed-in views of Fig. 8a are provided in Figs. 8b-8d. The raw images and reconstructed birefringence maps are presented in the first two images are overlaid in the last column of Fig. 8b-8d. The presence of hemozoin crystals inside infected cells allows for clear identification, enabling the visualization of the structure and birefringence of malaria-infected blood cells in high resolution. This capability of the pol-CP platform provides a promising avenue for a wide range of biomedicine-related applications, benefiting from its high-sensitivity birefringence imaging capability, as well as its ability to deliver high-resolution and large field-of-view imaging.

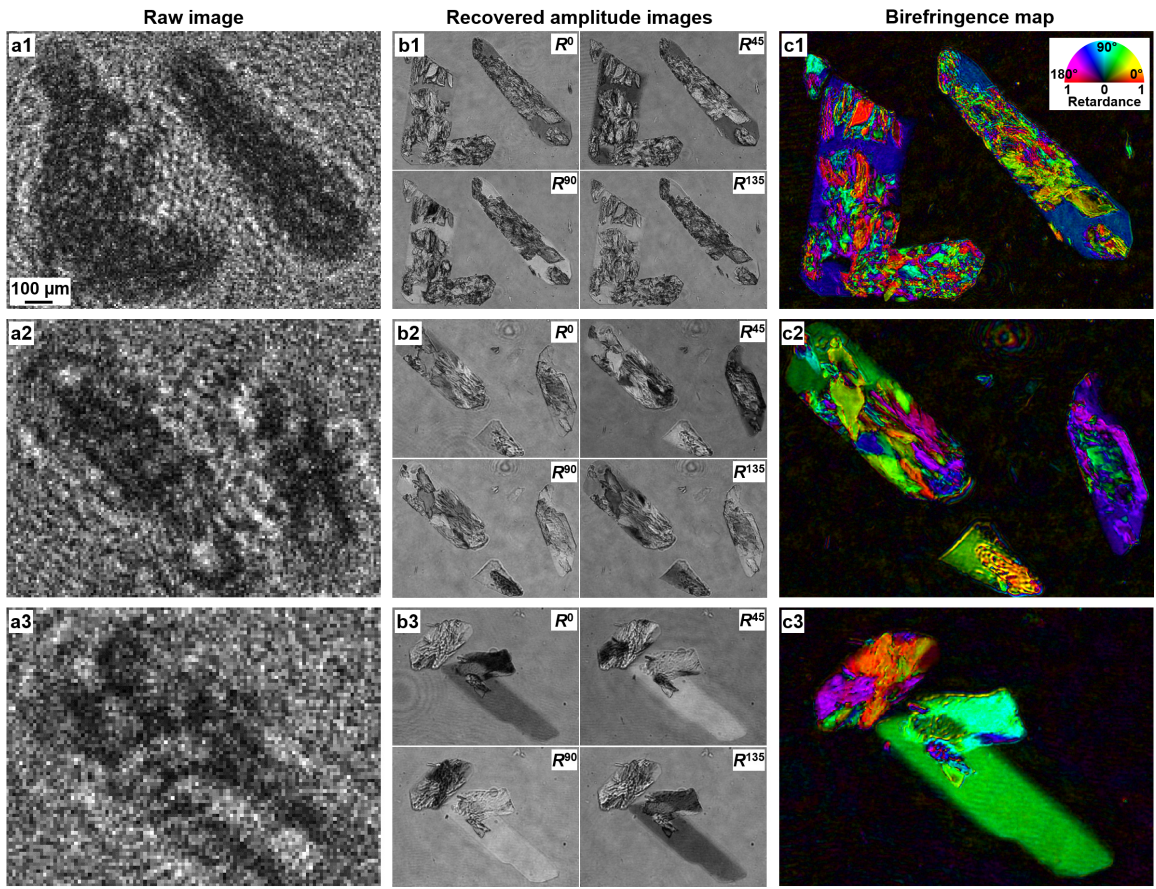

**Fig. 9.** Imaging of calcium phosphate crystal prepared from urine samples. (a1)-(a3) The captured raw image. (b1)-(b3) The recovered amplitude images at four polarization angles. (c1)-(c3) The recovered birefringence maps.

### 3.5 Polarimetric imaging of crystals

Crystal imaging plays a vital role across various scientific and industrial domains, such as materials science, pharmaceuticals, and geology. The ability to visualize and analyze crystals with quantitative polarimetric contrast offers profound insights into their physical attributes, chemical composition, and multifaceted applications. One prominent example lies in the medical field's analysis of urine samples. Urine samples often contain diverse crystal types, including calcium phosphate, calcium oxalate, uric acid, and cystine. These are influenced by factors such as an individual's diet, hydration level, and underlying health conditions. The presence and classification of these crystals can be symptomatic of specific diseases or disorders. For example, calcium oxalate crystals may indicate a susceptibility to kidney stones, uric acid crystals might suggest gout or metabolic disorders, and cystine crystals could reveal a rare genetic condition known as cystinuria, leading to cystine stones in the kidneys, bladder, and urethra. The ability to accurately image and analyze these crystals over a large field of view and with polarimetric contrast can provide valuable diagnostic information. It enables healthcare providers to identify the specific type of crystals present, assess their concentration, and understand their formation conditions. This information can guide treatment decisions, such as dietary modifications, medication adjustments, or further diagnostic testing.

To demonstrate the capability in crystal structure visualization, we imaged a sediment slide prepared from urine samples and identified calcium phosphate crystals in Fig. 9. The captured raw images of the crystals are shown in Fig. 9a while the recovered intensity images at four distinct polarization angles are shown in Fig. 9b. Figure 9c reveals the recovered birefringence maps of the calcium phosphate crystals. Another key advantage of the reported platform resides in its ability to perform digital refocusing after the data has been acquired. We substantiated this aspect using a calcium phosphate crystal sample in Fig. S3. Figures S3a-S3c illustrate the recovered object amplitude after digitally propagating to z = 1215 µm, 1229 µm, and 1240 µm, respectively. Visualization 2 further shows this propagation process.

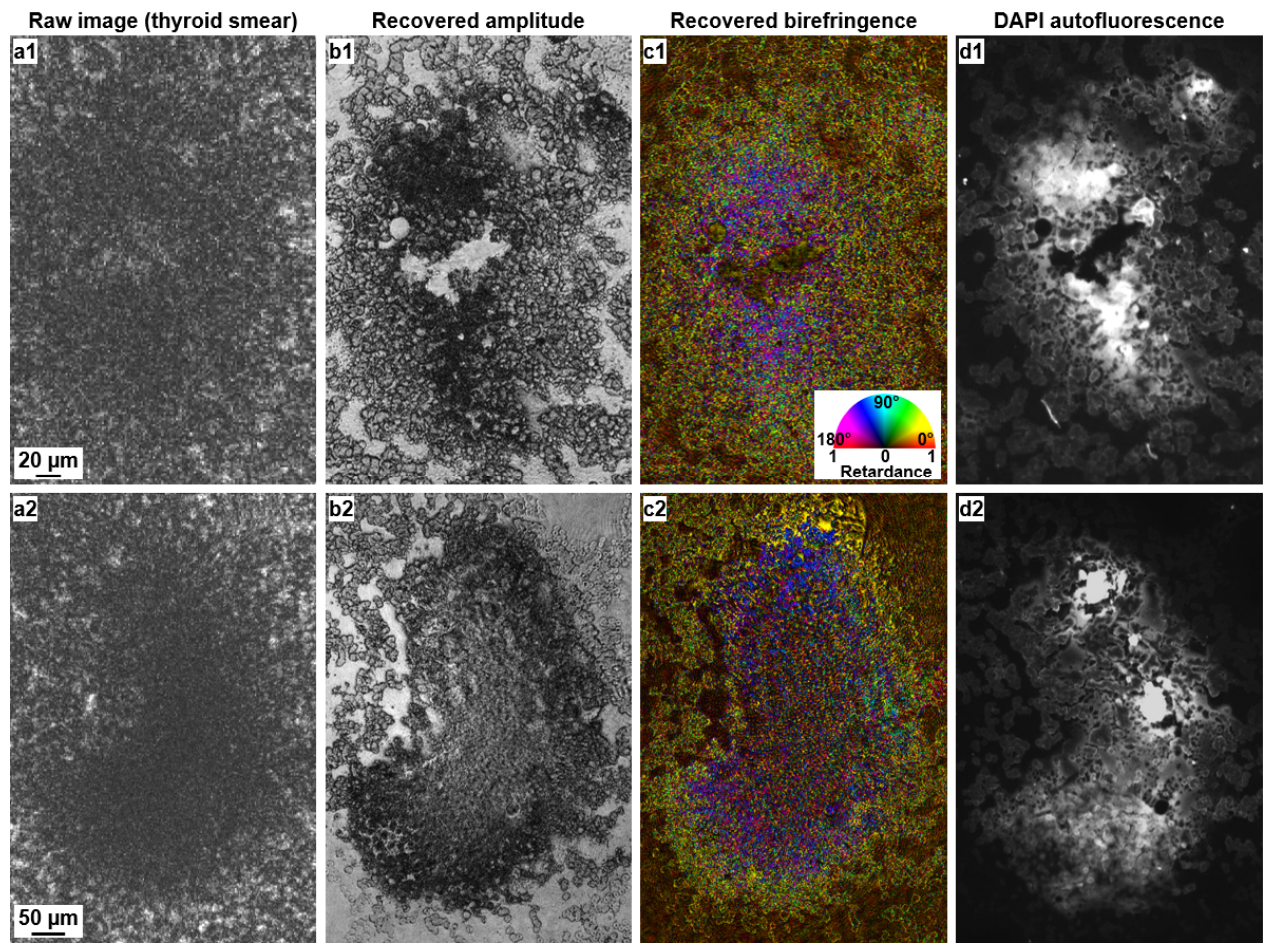

**Fig. 10.** Imaging of an unstained thyroid sample using the lensless pol-CP prototype. (a) The captured raw image. (b) The recovered amplitude. (c) The recovered birefringence map, where the calcium oxalate crystals are present within the colloid of thyroid follicles and contribute to the purple colour in the birefringence map. (d) The captured autofluorescence image captured using a regular fluorescence microscope with a DAPI filter cube. The bright region indicates the autofluorescence signals from calcium oxalates. The results demonstrated here can be used to locate follicles on unstained cytology smears for rapid on-site evaluation.

**3.6 Label-free chemical imaging of cytological smears**

Tissue samples are naturally transparent under a conventional light microscope. Staining is commonly employed to enhance visibility by adding colour to specific structures or features within a sample. While staining can aid in differentiation between different cell types or tissues, it is time-consuming for rapid on-site diagnostics and can potentially damage delicate structures. In contrast, our platform can provide label-free, intrinsic contrast by utilizing the polarization properties of biological specimens. To highlight this, we imaged an unstained thyroid smear using our prototype device in Fig. 10. Figures 10a-10d show the captured raw image, the recovered amplitude, the recovered birefringence map, and the DAPI autofluorescence map of the unstained thyroid smear, respectively. In particular, Fig. 10c shows differential contrast arising from calcium oxalate crystals [45]. We note that these crystals are present within the colloid of normal thyroid follicles, and they contribute to the birefringence properties shown in purple colour in Fig. 10c. Interestingly, the autofluorescence image captured using a regular fluorescence microscope with a DAPI filter cube highlights the same areas. This is consistent with the autofluorescence emission signal in a similar range from calcium oxalates [46]. The results demonstrated here can be used to locate follicles on unstained cytology smears for rapid on-site evaluation.

## 4. Discussion and conclusion

In this study, we introduced polarimetric coded ptychography (pol-CP) as a solution to the existing challenges of conventional lens-based polarization microscopy, overcoming the trade-off between spatial resolution and imaging field of view, as well as the constraints in cost, spatial resolution, field of view, and imaging throughput compared to robotic microscope platforms. Our work bridges the gap between the current limitations of conventional polarimetric imaging and the need for high-throughput, high-resolution imaging systems in biomedicine and beyond.

Through the integration of a coded sensor and an ultra-thin polarizer film, we achieved the acquisition of gigapixel birefringence images with a 180-mm$^2$ field of view in a time frame comparable to regular whole slide scanners. Notably, this was achieved in a cost-effective and portable manner without using any lens, rendering the platform highly suitable for field and resource-limited applications. The prototype platform also excels in its ability to refocus images after data acquisition, effectively addressing the autofocusing issue associated with traditional whole slide scanners. This ability enhances image quality, streamlines workflow, and minimizes the need for rescanning due to autofocusing errors. The elimination of the need to model the complex wavefront entering the sample, a constraint present in other methods such as FPM and illumination-based spatial-domain ptychography, increases flexibility, particularly in working with thicker specimens [22].

In practical applications, the pol-CP platform exhibited potential in malaria diagnosis by effectively tracking and detecting malaria-infected red blood cells. The birefringence changes associated with the infection process, particularly the formation of hemozoin crystals by the parasite, were effectively identified and visualized using our platform. Moreover, we demonstrated pol-CP's versatility in the realm of crystal imaging, particularly in the medical diagnosis of conditions indicated by the presence of specific crystals in urine samples. The platform's ability to provide detailed structural information through birefringence mapping offers insights into the type, concentration, and formation conditions of these crystals, thereby aiding in more accurate and timely treatment decisions. Our platform's potential was also highlighted in the label-free chemical imaging of unstained thyroid smears. Leveraging the inherent polarization properties of biological specimens, our platform offered intrinsic contrast, mitigating the need for time-consuming and potentially damaging staining processes. This ability can significantly enhance the efficiency and preservation of sample integrity in diagnostic processes, providing advancements for on-site rapid evaluations.

In conclusion, the pol-CP technique exhibits promise for the future of polarimetric imaging in the biomedical field. Its high-throughput imaging, high spatial resolution, low cost, portability, and flexibility with sample thickness anticipate its valuable role in disease diagnosis, sample screening, and label-free chemical imaging. We look forward to the continued development and application of this technique, bearing in mind that further work is required for its refinement, standardization, and exploration of utility in a broader range of biomedical applications.